\documentclass[twocolumn,tighten]{aastex631}
\hypersetup{linkcolor=red,citecolor=blue,filecolor=cyan,urlcolor=magenta}
\usepackage{amsmath}
\usepackage{natbib}
\usepackage{todonotes}
\usepackage{CJK}
\usepackage{xspace}

\newcommand{\Gaia}{\textit{Gaia}}

\newcommand{\kmsec}{\mbox{km~s$^{\rm -1}$}}

\newcommand{\msun}{\mbox{$M_{\odot}$}}

\newcommand{\RN}[1]{%
  \textup{\uppercase\expandafter{\romannumeral#1}}%
}
\newcommand{\rpro}{\mbox{{\it r}-process}\xspace}
\newcommand{\Rpro}{\mbox{{\it R}-Process}\xspace}
\newcommand{\spro}{\mbox{{\it s}-process}\xspace}

\newcommand{\teff}{\mbox{$T_{\rm eff}$}}
\newcommand{\logg}{\mbox{log~{\it g}}}
\newcommand{\vt}{\mbox{$v_{\rm t}$}}
\newcommand{\mh}{\mbox{[M/H]}}

\shorttitle{The Rise of the \rpro\ in Gaia-Sausage/Enceladus}
\shortauthors{X. Ou et al.}
\begin{document}
\begin{CJK*}{UTF8}{gbsn}

\title{The Rise of the \Rpro\ in the Gaia-Sausage/Enceladus Dwarf Galaxy}

\author[0000-0002-4669-9967]{Xiaowei Ou (欧筱葳)} 
\affiliation{%
Department of Physics and MIT Kavli Institute for Astrophysics and Space Research, \\ 
77 Massachusetts Avenue, Cambridge, MA 02139, USA}
\email{Email:\ xwou@mit.edu}

\author[0000-0002-4863-8842]{Alexander P. Ji}
\affiliation{Department of Astronomy \& Astrophysics, 
University of Chicago, 
5640 S. Ellis Avenue, Chicago, IL 60637, USA}
\affiliation{Kavli Institute for Cosmological Physics, 
University of Chicago, 
Chicago, IL 60637, USA}

\author[0000-0002-2139-7145]{Anna Frebel}
\affiliation{Department of Physics and MIT Kavli Institute for Astrophysics and Space Research, \\ 
77 Massachusetts Avenue, Cambridge, MA 02139, USA}

\author[0000-0003-3997-5705]{Rohan P. Naidu}
\altaffiliation{NASA Hubble Fellow}
\affiliation{Department of Physics and MIT Kavli Institute for Astrophysics and Space Research, \\ 
77 Massachusetts Avenue, Cambridge, MA 02139, USA}

\author[0000-0002-9269-8287]{Guilherme Limberg}
\affiliation{Universidade de S\~ao Paulo, Instituto de Astronomia, Geof\'isica e Ci\^encias Atmosf\'ericas, Departamento de Astronomia, \\ SP 05508-090, S\~ao Paulo, Brasil}

\begin{abstract}

Neutron star mergers (NSMs) produce copious amounts of heavy \rpro\ elements after a time delayed inspiral process. Once NSMs are present in a galaxy, \rpro elements, such as Eu, are expected to significantly increase with time.
Yet, there has been limited observational data in support of Eu increasing within Local Group galaxies. 
We have obtained high-resolution Magellan/MIKE observations of 43 metal-poor stars in the Gaia-Sausage/Enceladus tidally disrupted galaxy with $-2.5 < \rm{[Fe/H]} < -1$. 
For the first time, we find a clear rise in [Eu/Mg] with increasing [Mg/H] within one galaxy.  
We use a chemical evolution model to study how such a rise can result from the interplay of prompt and delayed \rpro\ enrichment events.
Delayed \rpro\ sources are required to explain the rise and subsequent leveling off of [Eu/Mg] in this disrupted galaxy.
% Prompt sources alone cannot produce the observed rise. 
However, the rise may be explained by delayed \rpro sources with either short ($\sim 10$\,Myr) or long ($\sim 500$\,Myr) minimum delay times.
Future studies on the nature of \rpro\ sources and their enrichment processes in the GSE will require additional stars in the GSE at even lower metallicities than the present study. 

\end{abstract}

% NEW from the UAT

\section{Introduction}
\label{sec:intro}

The origin of rapid neutron-capture process (\rpro) elements in the universe remains one of the most complex challenges in the realm of understanding the nucleosynthesis of all the elements across the periodic table \citep[see e.g.,][]{frebel18, cowan21}. The theoretical foundations of \rpro\ nucleosynthesis have been known for several decades \citep{burbidge57}, but the details are still not all fully understood.
The binary neutron star merger (NSM) event GW170817 provided the first direct evidence for NSMs as an astrophysical site for the operation of the \rpro\ \citep[e.g.,][]{drout17, kasen17, kasliwal17}.
Yet, it has been pointed out that NSMs may only be important at late(r) times due to delays induced by their binary evolution \citep[e.g.][]{cote18,vigna-gomez18,beniamini19,simonetti19,lian23}.
Rare types of core-collapse supernovae (rCCSNe) have been proposed as prompt \rpro\ sources to account for early \rpro\ production \citep[e.g.][]{mosta18,siegel19,kobayashi20,yong21} that appears to be required by observations of the many ancient metal-poor \rpro\-enhanced stars \citep{ezzeddine20,holmbeck20,shah24}.

Considering the chemical evolution of \rpro\ elements within a galaxy, the differences in delay time distribution between prompt and delayed sources predict a potentially observable increase in \rpro\ enhancement level as a function of time \citep{Matteucci2014,cescutti15,duggan18,skuladottir19}.
If all \rpro\ sources were prompt, the \rpro\ element abundances, such as Eu, are expected to be constant relative to the $\alpha$-element abundances, e.g. Mg, that are produced by prompt core-collapse supernovae (CCSNe).
Delayed \rpro\ sources, however, would cause the ratio of Eu to Mg abundances to increase as time passes/star formation progresses.
While a comparison of [Eu/Mg] ratios in different galaxies suggested delayed \rpro\ to play an important role \citep{skuladottir20,reggiani21,matsuno21,naidu22}, clear evidence of a rise in [Eu/Mg] as a function of metallicity within a single galaxy has been lacking.
Instead, constant [Eu/Mg] ratios are observed in most local group galaxies, hinting that prompt sources could account for all \rpro\ production (e.g., \citealt{skuladottir19,reichert20}), with the recent exception of Wukong \citep{limberg23} and possibly also Ursa Minor \citep{duggan18}.

Gaining a better understanding of \rpro\ sources and enrichment processes becomes possible with measurements of \rpro\ abundances in stars that (1) span a significant range of metallicities, and 
(2) are associated with a common birth environment that we can clearly identify and characterize. 
These two conditions are both essential for establishing evidence of delayed sources but are challenging to achieve.  
For example, in the Milky Way, it is difficult to properly trace the galactic \rpro\ content back to its sources because of its complex merger history \citep[e.g.,][]{tsujimoto14,ishimaru15,ji16}, but stars at nearly all metallicities are accessible.
However, efforts are underway from spectroscopic surveys such as the H3 survey \citep{conroy19} to map out the merger history and trace the galactic \rpro\ sources.
On the other hand, for any intact dwarf galaxies, measurements of \rpro\ abundances spanning a significant range in metallicities have proven challenging to obtain due to their distances ($\gtrsim 100$\,kpc) and corresponding faintness, and general low stellar mass ($\lesssim 10^6$\,\msun) \citep{frebel23}, although recent progress has been made with the Magellanic clouds at $\sim 50$\,kpc \citep{reggiani21,oh24,chiti24}.

To circumvent these challenges, recent studies have focused on numerous stars from merger debris of disrupted dwarfs ($\gtrsim 10^7$\,\msun) now found across the local halo of the Milky Way.
In particular, the Gaia-Sausage/Enceladus (GSE) ($M_{\star} \sim 10^{8.5}$-$10^{9.5}$\,\msun; \citealt{haywood18,helmi18,belokurov18,naidu20,kruijssen20,limberg22}) is one such system that is expected to have had a less complex merger history prior to merging with the Milky Way. 
The distinct, highly radial kinematics (orbital eccentricity $e \gtrsim 0.7$) and relatively close distances of the GSE stars allow their efficient identification for obtaining high-resolution spectroscopy. 
In addition, it is the Milky Way's most massive accretion event, with star formation lasting long enough ($\sim3.6$\,Gyr; \citealt{bonaca20}) to have likely experienced \rpro\ enrichment from delayed sources.

% The stellar mass of the GSE progenitor is estimated to be $\sim 10^8$-$10^9$\,\msun, with an accretion time $\sim10$-$12$\,Gyr ago \citep{haywood18,helmi18,belokurov18,bonaca20,naidu20,kruijssen20}.
% Due to its relatively massive stellar component and radial infall, our local ex-situ stellar halo is dominated by stars originating from the GSE progenitor \citep{naidu20}. 
% The distinct, highly radial kinematics (orbital eccentricity $e \gtrsim 0.7$) of the GSE stars thus allow their efficient identification. Given their relatively close distances, high-resolution spectroscopic observations of these stars become feasible.

Accordingly, a number of studies based on low and high-resolution spectra of GSE stars have started to map out the chemical evolution of the GSE progenitor. 
% Estimates of the average metallicity of the GSE range from [Fe/H]$\sim -1.6$ to $-1.1$ \citep{helmi18,naidu20}.
% \citet{hasselquist21} found that the GSE was $\alpha$-element enhanced up to [Fe/H] $\sim -1.2$, suggesting an efficient early star formation.
% Deficiencies observed in [Ce/Mg] and [(C+N)/Mg], on the other hand, indicate that star formation was likely quickly quenched upon the merger with the Milky Way.
Recent studies have focused on \rpro\ elements, finding that GSE stars exhibit an overall enhancement in Eu abundances \citep{matsuno21,aguado21b,dasilva23}. 
\citet{naidu22} compared [Eu/Mg] ratios in both the GSE and Kraken stars \citep{kruijssen20}, two systems with similar stellar mass but different star formation duration. They argued that both rCCSNe and NSMs\footnote{Following \citet{naidu22}, we will refer to prompt (delayed) sources and rCCSNe (NSMs) interchangeably hereafter.} are needed to explain the [Eu/Mg] abundances observed in the GSE population. 

Previous studies, however, have failed to probe the early \rpro\ enrichment process in the GSE, as samples had been restricted to $\rm{[Fe/H]}\gtrsim-2.0$. 
As pointed out by \citet{naidu22}, only stars with lower metallicity ($\rm{[Fe/H]}<-2$) would have formed from gas dominantly enriched by prompt sources, such as rCCSNe.

In this study, we combine a new sample of metal-poor GSE stars with the existing GSE sample studied in \citet{naidu22} to extend the metallicity coverage to $-2.5<\rm{[Fe/H]}<-1.0$.
With 43 GSE stars in total, we consistently re-derive stellar parameters and chemical abundances with a particular focus on Eu and Mg, using Eu as a proxy for \rpro\ enrichment and Mg as a proxy for the occurrence rate of CCSNe (Section~\ref{sec:target_selection_obs}).
A clear rising trend in [Eu/Mg] abundances with metallicity is observed for the first time in the GSE (Section~\ref{sec:result}), confirming the existence of delayed \rpro\ sources. 
% We then translate the [Mg/H] abundances into the ages of the stars using an empirically constructed relation for the GSE to compare our results with a simple chemical evolution model that includes prompt and delayed \rpro\ enrichment.
We qualitatively explore the rate at which Eu is produced with a simple chemical evolution model invoking prompt and delayed sources and their different timescales (Section~\ref{sec:discussion}) based on the observed abundance trends. 
Such a rise is not observed in most other dwarfs (Section~\ref{sec:compare}).
Our results thus show that NSMs must have been present as a dominant site of \rpro\ production in the GSE (Section~\ref{sec:conclusion}). 

\section{Observations and Analysis}
\label{sec:target_selection_obs}

\begin{deluxetable*}{cccccccccc}
\tablecaption{Stellar parameters, final derived chemical abundances, and observing information for 43 stars considered in this study. The uncertainties listed include statistical and systematic uncertainties.}
\label{tab:sumtab}
\tablehead{
\colhead{\Gaia\ DR3 Source ID} & \colhead{\teff} & \colhead{\logg} & \colhead{[Fe/H]} & \colhead{$\sigma_\text{[Fe/H]}$} & \colhead{[Mg~\textsc{i}/H]} & \colhead{$\sigma_\text{[Mg~\textsc{i}/H]}$} & \colhead{[Eu~\textsc{ii}/H]} & \colhead{$\sigma_\text{[Eu~\textsc{ii}/H]}$} & \colhead{[Eu~\textsc{ii}/Mg~\textsc{i}]}
}
\startdata
31707926776039680 & 4896 & 1.89 & $-$1.96 & 0.18 & $-$1.63 & 0.12 & $-$1.57 & 0.09 & 0.06 \\
1187898485610541696 & 4451 & 1.53 & $-$1.19 & 0.18 & $-$0.95 & 0.14 & $-$0.42 & 0.12 & 0.52 \\
2472646349146818816 & 4593 & 1.51 & $-$1.55 & 0.21 & $-$1.21 & 0.12 & $-$0.83 & 0.11 & 0.38 \\
2504634406573540608 & 4814 & 2.15 & $-$1.24 & 0.21 & $-$1.00 & 0.14 & $-$0.72 & 0.12 & 0.27 \\
2505065101598396416 & 4488 & 1.49 & $-$1.32 & 0.19 & $-$1.03 & 0.14 & $-$0.74 & 0.10 & 0.29 % \\
\enddata
\tablecomments{%
The complete version of Table~\ref{tab:sumtab} is available in the online edition of the journal. 
A short version with a subset of the columns is included here to demonstrate its form and content.
The full table contains the following columns: \Gaia\ DR3 Source ID, coordinates, observing details (date, binning, exposure time, and slit size), \teff, \logg, [Fe/H], \vt, abundances ([X/H]) for Mg, Ba, Eu, and their associated uncertainties.
}
\end{deluxetable*}

The metal-poor GSE stars are selected following the same procedure described in \citet{naidu22}.
Briefly, stars are selected from APOGEE DR16 \citep{jonsson20}, the H3 survey \citep{conroy19}, and the Best \& Brightest selection \citep{schlaufman14,placco19,limberg21b} crossmatched with \Gaia\ DR3 \citep{gaia21}. 
Spectroscopic surveys provide line-of-sight velocities, while \Gaia\ provides the coordinates, parallaxes, and proper motions.
Dynamical quantities are then computed according to \citet{naidu20} using the default \texttt{MilkyWayPotential}~from \textsc{gala} Python package \citep{gala_software}.

We select GSE stars from APOGEE DR16 in the chemodynamical space using the following criteria: 

\begin{equation}
\begin{aligned}
(r_{\rm{gal}}/[\rm{kpc}]>5) \land\ (\textit{e}>0.7)\\
\land\ \rm{[Mg/Mn]}>0.25\\
\land\ \rm{[Mg/Mn]}-4.25\rm{[Al/Fe]}>0.55\\
\land\ E_{\rm{tot}}/[10^{5}\ \rm{km}^{2}\ \rm{s}^{-2}]>-1.50.
\end{aligned}
\end{equation}

The [Mg/Mn] and [Al/Fe] space isolates the GSE accreted stars from the in-situ Milky Way stars \citep{hawkins15}.
The dynamical space selections are similar to those in the literature \citep[e.g.,][]{limberg22,buder22}.

For stars from the H3 survey, we select GSE stars as described in \citet{naidu20}. For the Best \& Brightest sample, we select GSE stars following \citet{feuillet20} using 

\begin{equation}
\begin{aligned}
|J_{\phi}|/[\rm{kpc}\ \rm{km}\ \rm{s}^{-1}]<500 \\
\land\ J_{r}/[\rm{kpc}\ \rm{km}\ \rm{s}^{-1}]>900 \\
\land\ J_{r}/[\rm{kpc}\ \rm{km}\ \rm{s}^{-1}]<2500.
\end{aligned}
\end{equation}

High-resolution optical spectra of 43 stars were obtained using the Magellan Clay $6.5$\,m Telescopes at Las Campanas Observatory using the Magellan Inamori Kyocera Echelle (MIKE) spectrograph \citep{bernstein03} from 2021 July to December. 
Eleven stars were observed with the 0\farcs5 slit and 2x1 binning in July 2021, yielding a resolution $R \sim 40{,}000/50{,}000$ on the blue/red arm of MIKE with wavelength coverage 3200--5000 and 4900--10000\,\AA, respectively.
The other 32 stars were observed with the 0\farcs7 slit and 2x2 binning with a resolution $R \sim 28{,}000/22{,}000$. 
Table~\ref{tab:sumtab} describes the observational date, exposure time, and instrument setup for each observed target.

Stellar parameters were derived photometrically. 
Effective temperature (\teff) was determined using Gaia DR3 $G-RP$ colors \citep{gaia21} using color-[Fe/H]-\teff\ relations from \citet{mucciarelli21}. 
Metallicity (\mh) was set to be the average of Fe\,\textsc{i} line abundances measured using ATLAS model atmospheres from \citet{castelli03}, with [$\alpha$/Fe] set equal to [Mg/Fe]. 
Surface gravity (\logg) was determined by interpolating the Dartmouth isochrones along the red giant branch \citep{dotter08}. 
A pure spectroscopic analysis confirmed that these stars were all giant stars.
Microturbulence (\vt) was set to maintain reduced equivalent width (REW) balance in Fe\,\textsc{i}. 
The derivation was iterated until the final photometric stellar parameters stabilized for each star. 

Statistical uncertainties on \teff~and \logg~were determined by propagating input uncertainties through the color-[Fe/H]-\teff\ relations and isochrone fitting, respectively. 
\mh~and \vt~statistical uncertainties were determined from the standard error in [Fe\,\textsc{i}/H] and the slope with respect to REW, respectively. 
Systematic uncertainties for \teff, \logg, \mh, and \vt~were set to 100~K, 0.2~dex, 0.2~dex, and 0.2~\kmsec. 
Final stellar parameters and associated total uncertainties for all $43$ stars are included in Table~\ref{tab:sumtab}. 

We adopted the line list from \citet{naidu22} with weak lines that are not measurable in metal-poor stars ($\text{[Fe/H]}<-1.5$) removed. 
Abundances were determined following \citet{naidu22}. 
The spectra were analyzed using \texttt{SMHR}\footnote{\url{https://github.com/andycasey/smhr}} \citep{casey14}, which provides an interface for fitting equivalent widths, interpolating the stellar atmosphere models from \citet{castelli03}, and running MOOG including scattering \citep{sneden73,sobeck11} and \citet{barklem00} damping to determine abundances from equivalent widths and spectrum synthesis. 
A more detailed analysis description can be found in \citet{ji20}. 
We show the final derived abundances in Table~\ref{tab:sumtab} and Figure~\ref{fig:eu_mg_feh}.

\section{Rise of the \textit{R}-process}
\label{sec:result}

We first examine the $\alpha$-element abundances [Mg/Fe] to further check on the GSE membership of the stars in our sample, in particular whether they are accreted or in situ interlopers. 
\citet{nissen10} identified two distinct halo populations in the solar neighborhood with high-$\alpha$ and low-$\alpha$ abundances and associated them with in-situ and accreted origin, respectively. 
We take these populations as a benchmark for our sample. 
Comparing our measured [Mg/Fe] abundance with those from \citet{nissen10} shows that they are consistent with their low-$\alpha$ population stars. 
% The top panel of Figure~\ref{fig:eu_mg_feh} shows that our sample's [Mg/Fe] values are broadly consistent with this low-$\alpha$ population, with $\text{[Mg/Fe]}\lesssim0.3$ at $\text{[Fe/H]}\gtrsim-1.5$. 
Furthermore, comparisons between our sample with other studies targeting GSE stars \citep{aguado21b, matsuno21} show overall good agreement. 
Thus, we confirm that the selection described in Section~\ref{sec:target_selection_obs} yielded a pure GSE sample.

\begin{figure}
    \centering
    \includegraphics[width=0.4\textwidth]{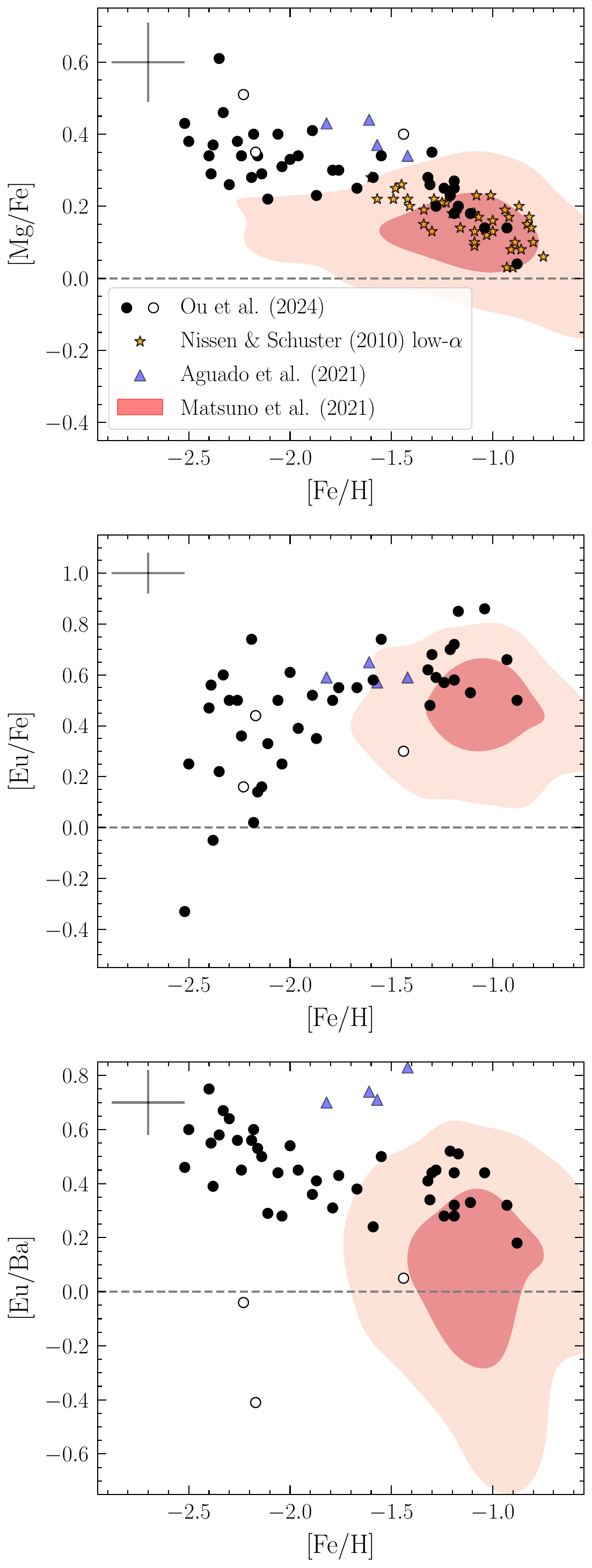}
    \caption{[Mg/Fe] (top), [Eu/Fe] (middle), and [Eu/Ba] (bottom) as a function of [Fe/H]. Measurements from this study are shown in black-filled/open circles. 
    The typical uncertainties in the measurements are shown as the grey error bar in each panel.
    The orange stars are the low-$\alpha$ stars from \citet{nissen10}. The blue triangles are the GSE stars from \citet{aguado21b}. The red contour shows the GSE stars from \citet{matsuno21}.}
    \label{fig:eu_mg_feh}
\end{figure}

In the middle panel of Figure~\ref{fig:eu_mg_feh}, we show the increase in [Eu/Fe] as a function of [Fe/H]. The metal-rich ($\text{[Fe/H]}\gtrsim-1.75$) part of the sample shows a consistently elevated mean value of $\text{[Eu/Fe]}=0.62 \pm 0.02$\,dex with a standard deviation $\sigma=0.13$\,dex. 
The same behavior was also found by \citet{aguado21b} and \citet{matsuno21}. 
The metal-poor sample ($\text{[Fe/H]}\lesssim-1.75$), on the other hand, exhibits a lower level of $\text{[Eu/Fe]}=0.39\pm0.01~(\sigma=0.22)$\,dex.

Most stars in our sample have large [Eu/Ba] ratios that indicate an \rpro\ origin for these elements, bottom panel of Figure~\ref{fig:eu_mg_feh}. Specifically, the \rpro nucleosynthesis predicts a ratio of $\text{[Eu/Ba]}=0.8$ for a pure \rpro\ origin \citep{sneden08}. Some of our stars fall near the $\text{[Eu/Ba]}=0.8$ regime, while others have slightly lower values (between $0.2$ and $0.6$). This suggests that some \spro\ contribution is present in these stars, though the \spro\ contribution to Eu is low enough to be neglected. Three additional stars show signatures of significant \spro\ contribution, given their low [Eu/Ba] values of $\text{[Eu/Ba]}\lesssim0.1$. We mark these stars as open circles in all Figures and ignore them for studying \rpro\ evolution.

% Regarding the stars with $\text{[Eu/Ba]}=0.2$ to $0.6$, we assess the level of \spro\ contamination. Assuming solar \rpro\ and \spro\ patterns \citep{sneden08}, the Eu contribution from \rpro\ is estimated to be $\sim150$ times that from \spro, in terms of Eu mass. This estimate is an over-simplification because the \spro\ pattern is somewhat metallicity dependent \citep[e.g.,][]{lugaro12}, but it nevertheless illustrates that the Eu in our GSE stars is predominantly produced in \rpro\ events. We thus conclude that overall, any contributions to Eu by the \spro\ are minimal in GSE.

\begin{figure*}
    \centering
    \includegraphics[width=\textwidth]{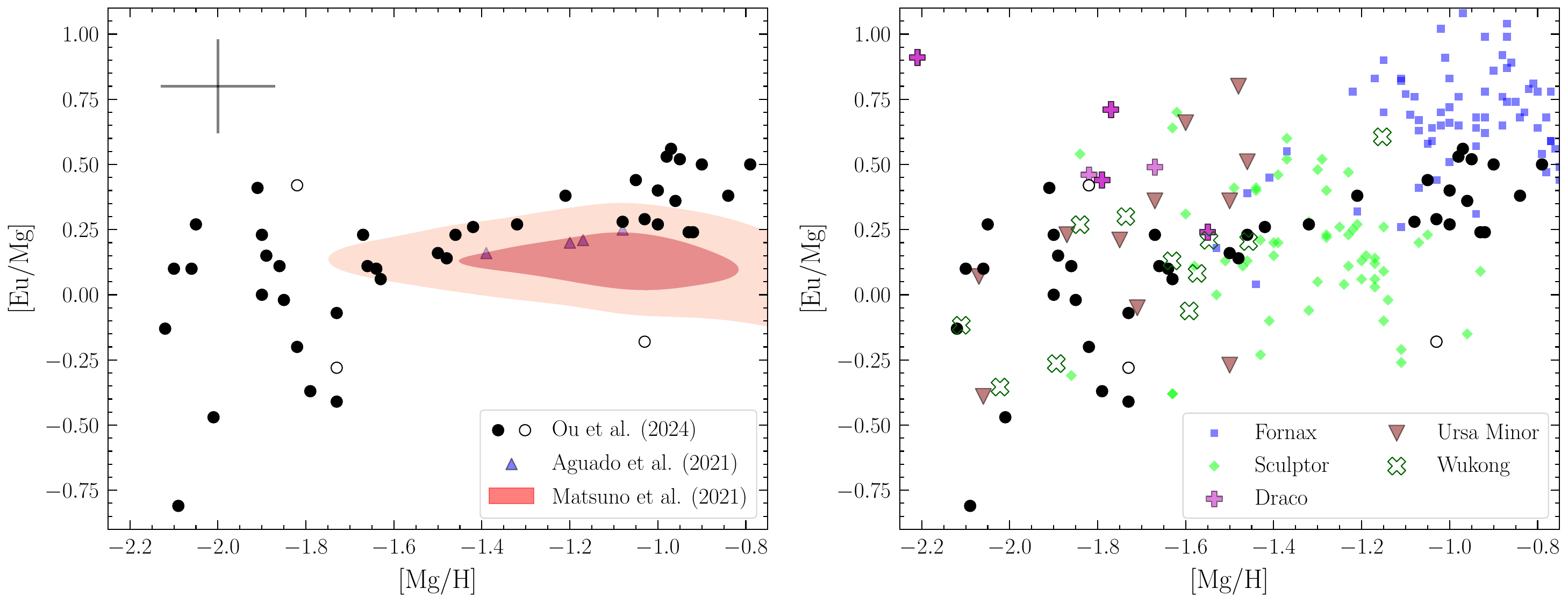}
    \caption{[Eu/Mg] as a function of [Mg/H].
    Measurements from this study are shown in black-filled/open circles. 
    The typical uncertainty of the measurements is shown as the grey error bar.
    The left panel compares our sample with other GSE \rpro\ measurements \citep{matsuno21,aguado21b}. 
    The right panel compares our sample with other dwarf systems, with intact dwarfs Fornax, Sculptor, Draco, and Ursa Minor (compiled and taken from \citealp[]{frebel23}), as well as the dissolved dwarf Wukong/LMS-1 sample \citep{limberg23}. 
    }
    \label{fig:eumg_mgh_comp}
\end{figure*}

In order to remove the effect of the delayed production of Fe by Type~Ia SNe, we choose to consider [Eu/Mg] as a function of [Mg/H] \citep[e.g.,][]{tolstoy09,skuladottir20,limberg23} going forward. The evolution of the [Eu/Mg] is shown in Figure~\ref{fig:eumg_mgh_comp}, and can be interpreted as the evolution of \rpro\ elements as a function of time and star formation.

We find an increase in the mean [Eu/Mg], with $\text{[Eu/Mg]}=0.05 \pm 0.03~(\sigma=0.24)$ and $\text{[Eu/Mg]}=0.32 \pm 0.04~(\sigma=0.16)$ for stars with $\text{[Mg/H]}<-1.5$ and $\text{[Mg/H]}>-1.5$, respectively.
Thus, for the first time, we observe a robust rise of [Eu/Mg] with decreasing scatter over the course of the star formation history in the GSE. 
The observed rising trend is strong evidence of delayed \rpro\ sources, presumably NSMs. 

\section{Modeling the \textit{R}-process rise in the GSE}
\label{sec:discussion}

\subsection{\textit{R}-process model with rCCSNe and NSMs}
\label{sec:discussion_setup}

We model the rise in [Eu/Mg] as the result of two dominant \rpro\ sites, prompt (rCCSNe) and delayed (NSMs) sources, and estimate the yields and delay time distribution necessary to reproduce the observed rise. 
To do so, we first build an empirical relation between [Mg/H] and age.
Assuming the total mass of Mg in the GSE increases linearly with time, [Mg/H] increases logarithmically with time, i.e., the age of the star decreases exponentially with [Mg/H].
Given the infall time of the GSE ($\sim10$\,Gyr ago; e.g., \citealt{bonaca20}) and our sample with $-2.2\lesssim\rm{[Mg/H]}\lesssim-0.8$, we construct the exponential relation:

\begin{equation}
\rm{Age} = -0.15 \times 10^{\rm{[Mg/H]}+2.2} + 14.
\end{equation}

\begin{figure*}
    \centering
    \includegraphics[width=0.8\textwidth]{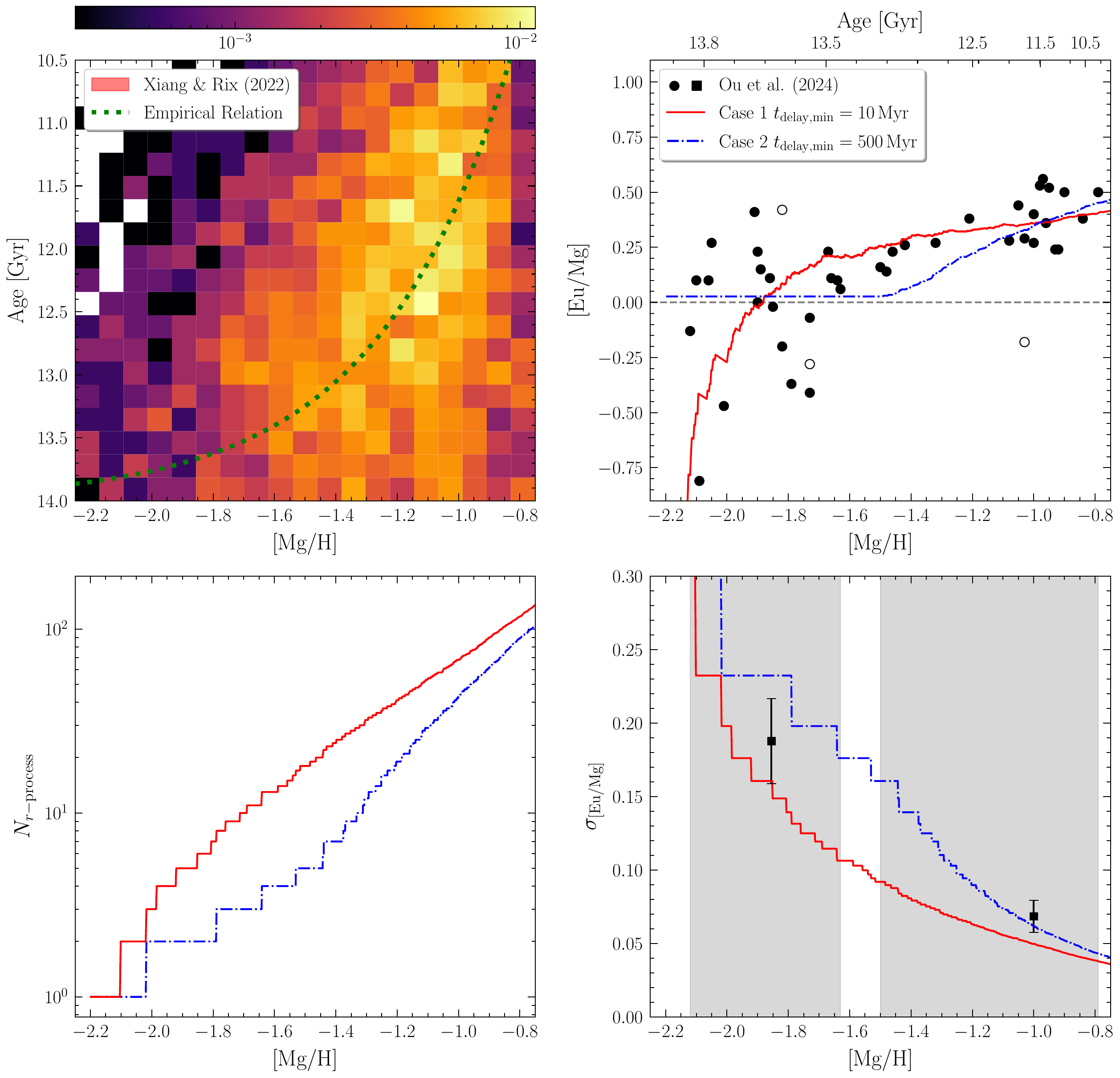}
    \caption{
    Comparison of our simple model with observational data. The model is further described in Section~\ref{sec:discussion}.
    The top left panel shows the empirical relation (green dashed line) between [$\alpha$/H] and age compared with GSE candidates from \citet{xiang22}.
    The 2D histogram shows the distribution of the stars in log scale.
    % normalized for each [Mg/H] bin.
    The top right panel shows two cases of our model (red solid line and blue dot-dashed line) of the Eu enrichment process by rCCSNe+NSMs.
    The bottom left panel shows the corresponding number of \rpro\ events for the two cases. 
    The bottom right panel shows the observed (squares) and expected (lines) scatters in [Eu/Mg] as a function of [Mg/H]. 
    The expected scatters are estimated from the number of \rpro\ events based on Poisson statistics, while the observed scatters (after removing the typical measurement uncertainties in [Eu/Mg]) are calculated from two sub-samples split up at $\text{[Mg/H]}=-1.5$. 
    The vertical error bars represent the estimated uncertainties of the observed scatter.
    The shaded regions represent the range of [Mg/H] of the two sub-samples, with the horizontal position of the squares marking the median [Mg/H]. 
    }
    \label{fig:eumg_mgh_age_model}
\end{figure*}

\citet{xiang22} obtained age estimates for subgiants by comparing the stars' locations to isochrones.
We thus select GSE candidates from their sample following our dynamical space selections described in Section~\ref{sec:target_selection_obs} for comparison. 
As shown in the top panels of Figure~\ref{fig:eumg_mgh_age_model}, assuming their [$\alpha$/H] is the same as our [Mg/H], our relation is within the scatter of ages from \citet{xiang22} and provides a reasonable description of GSE's chemical enrichment history. 
We note that this conversion between [Mg/H] and age is intended only for estimating the timescale of the Eu abundance rise, and not for actually assigning ages to our GSE stars.
Thus, stars with ages older than the Hubble time do not affect our ability to interpret the time evolution of \rpro\ in the GSE. 

Our chemical evolution model is the same as the one introduced by \citet{naidu22}. 
In short, the model describes the trend with rCCSNe, a prompt \rpro\ site, and NSMs, a delayed \rpro\ site producing Eu, as well as normal CCSNe producing Mg in a galaxy with a constant star formation rate. 
The normal CCSNe rate is set to produce [Mg/H] evolution consistent with the empirical relation between stellar age and [Mg/H].
For the prompt site, we assume a constant production rate of Eu as a function of time with an effective yield per one Myr of star formation ($y_{\rm{eff,p}}$). 
For the delayed site, we assume a power law delayed time distribution ($t_{\rm{delay}} \propto t^{-1.5}$; e.g., \citealt{paterson20,zevin22})\footnote{The power law exponent choice between $-1.5$ and $-1$ has no significant impact on the conclusion of the analysis.} with a minimum delay time ($t_{\rm{delay,min}}$) for the Eu production with effective yield per one Myr of star formation ($y_{\rm{eff,d}}$). 
$y_{\rm{eff,p}}$ and $y_{\rm{eff,d}}$ are determined both by the relative rate of rCCSNe (or NSMs) to CCSNe and the yield of individual rCCSNe (or NSMs).
We thus have a total of three free parameters for the model: $y_{\rm{eff,p}}$, $y_{\rm{eff,d}}$, and $t_{\rm{delay,min}}$.
Due to the degeneracies discussed below, in practice, we fix $t_{\rm{delay,min}}$ and set the effective yields to match the model to the data.

\subsection{Two cases and inferred yields}
\label{sec:discussion_result}

This model can reasonably describe the observed [Eu/Mg] trend with both short ($10$\,Myr) and long ($500$\,Myr) minimum delay time for NSMs, as shown in Figure~\ref{fig:eumg_mgh_age_model}, given the large scatter in [Eu/Mg] at ages older than 13\,Gyr.
With the short minimum delay time (Case 1; red line in the top right panel of Figure~\ref{fig:eumg_mgh_age_model}), we interpret the $\alpha$-poor sample as a fast increase in [Eu/Mg] starting from the most $\alpha$-poor stars, with a decreasing slope as age decreases. 
With the long minimum delay time (Case 2; blue line), we interpret the $\alpha$-poor sample to be constant with a large scatter up to age $13$\,Gyr, as expected with only rCCSNe. 
In this case, [Eu/Mg] starts increasing after the NSMs have experienced the minimum delay time.

We find that the effective Eu yield for NSM is qualitatively the same in both cases. 
This is expected as the two cases are constrained to reach the same mean observed level of [Eu/Mg] at $\mbox{[Mg/H]} > -1.2$.
At this [Mg/H] range, the Eu production is dominated by the NSM in either case, and thus the total Eu yield for these NSMs must be similar. 
Assuming a relative rate of NSM to CCSNe of $10^{-2.5}$, based on estimates for NSMs from gravitational wave observations \citep{ligo21} and CCSNe from transient surveys \citep{perley20}, we get a Eu yield per NSM of $10^{-5.5}$\,\msun, consistent with estimates from GW~170817 \citep{cote18} and LIGO-Virgo-KAGRA NSM observations \citep{chen24}.

The biggest difference between the two cases is the Eu yield from rCCSNe. 
For Case 1, the short minimum delay scenario, we find that a much lower effective Eu yield from rCCSNe is required than the long minimum delay scenario model (Case 2).
In other words, the rCCSNe do not need to produce nearly as much Eu if the NSMs start producing Eu with a short time delay.
This is consistent with the fact that rCCSNe are invoked to provide early \rpro\ production if NSMs experience long delay time before their merger (e.g., \citealt{Matteucci2014,siegel19}).
Assuming a relative rate of rCCSNe to CCSNe of $10^{-4}$, consistent e.g. with long gamma-ray burst rates \citep{guetta05,wanderman10,lien14}, the Eu yield per rCCSNe is $\sim5\times10^{-7}$\,\msun\ and $\sim1\times10^{-5}$\,\msun\, in Case 1 and Case 2, respectively. 
The latter is consistent with results from \citet{siegel19} and \citet{brauer21}, as expected for cases where collapsars can produce the Eu abundances observed in very metal-poor stars. 

We emphasize again that the increasing [Eu/Mg] requires delayed \rpro\ sources, regardless of the minimum delay time. 
Furthermore, as long as the effective Eu yield for NSM is fixed to match the final observed [Eu/Mg], the minimum delay time of NSMs can be reduced by decreasing the effective Eu yield for rCCSNe.
The effective Eu yield of the rCCSNe is, therefore, degenerate with the minimum delay time of NSMs.
Thus, fast mergers from NSMs mimic early rCCSNe, rendering the two indistinguishable from each other, as it may have occurred in the GSE.

\subsection{Breaking the degeneracy}
\label{sec:discussion_break_deg}

The degeneracy in the minimum delay time can be broken with \rpro\ measurements from other galaxies.
\citet{naidu22} showed that the minimum delay time must be $\gtrsim 500$\,Myr by comparing the median [Eu/Mg] of stars from the GSE and Kraken, two disrupted dwarfs with a similar stellar mass but different stellar formation duration.
We note that this alternative method relies on assumptions about the galaxies' star formation history and accretion time, and can only place a lower limit on the minimum delay time.
However, the star formation history and accretion time for Kraken are relatively uncertain.
Nonetheless, we may tentatively argue that Case 2 with a 500\,Myr delay time for the NSMs and a higher rCCSNe \rpro yield is preferred when taking into account the constraint from \citet{naidu22}.

The scatter in [Eu/Mg] can also break the degeneracy.
The intrinsic rare nature of \rpro\ events means that there must be a transition from early discrete stochastic \textit{enrichment} ($\sim$a few events) to a late continuous \textit{evolution} ($\sim$30 events) \citep{brauer21,frebel23}. 
While the onset of any delayed \rpro\ sources increases the mean \rpro\ element abundances, the overall transition from chemical enrichment to evolution tends to reduce the scatter around the mean. 
The bottom left panel of Figure~\ref{fig:eumg_mgh_age_model} shows the expected total number of \rpro\ events in the two cases. 

We infer the intrinsic scatter in [Eu/Mg] based on Poisson statistics from the number of events, assuming no intrinsic variation in \rpro\ events yield and there is instantaneous mixing. 
% We find $\sim0.1$\,dex for Case 1 and $\sim0.2$\,dex for Case 2. 
After accounting for typical measurement uncertainties ($\sim0.15$\,dex), the observed scatter in [Eu/Mg] is qualitatively consistent with the values we found for either case, as shown in the bottom right panel of Figure~\ref{fig:eumg_mgh_age_model}.
Further quantifying and constraining the total number of \rpro\ events via the observed scatter in [Eu/Mg], however, requires a larger sample of stars at low [Mg/H] to improve the statistics. 
Such a sample could be obtained for the GSE, given that most of the local halo is made of GSE debris. 

\section{\textit{R}-process in other dwarf galaxies}
\label{sec:compare}

We now consider available \rpro\ abundances in intact and disrupted dwarf galaxies to establish to what extent other systems might also display the steep rise in [Eu/Mg] observed in the GSE.
%xx(AJi: should this just be a different section? It is very distinct. The part below this is overall a bit confusing and not clear what is the point of the discussion, needs some work.) (XO: I made this a separate section, moved its place, and condensed the information to highlight the rise missing.)

In Figure~\ref{fig:eumg_mgh_comp}, we compare \rpro\ abundance measurements available for stars in other dwarf galaxies with our sample. 
For the most massive systems (stellar mass $M_\star > 10^6\,\msun$ from \citealp[]{mcconnachie12}), Sculptor and Fornax, the stars roughly overlap with the GSE sample at high [Mg/H].
Specifically, for Sculptor, measurements of the Eu abundances from \citet{hill19} established that the system has reached a constant $\text{[Eu/Mg]}\sim0.2$ at higher metallicities. 
Yet, the existence of delayed \rpro\ sources in Sculptor is still under debate because stars with actual Eu measurements remain lacking at $\text{[Mg/H]}\lesssim-1.6$ \citep{duggan18,skuladottir19}. 
% This prevents any strong conclusions regarding the evolution of [Eu/Mg] for Sculptor.

Likewise, Fornax \citep{letarte10,lemasle14} shows a constant elevated level of $\text{[Eu/Mg]}\sim0.4$, with no conclusive evidence of a change in [Eu/Fe]. Hence, there is no support for delayed sources in Fornax due to a lack of Eu measurements at lower metallicities.

On the other hand, the medium-mass ($M_\star \sim 10^{5.5}\,\msun$) dwarf galaxies, Draco and Ursa Minor, show more significant scatter and no clear trend compared to what we observed in the GSE.
The large scatter suggests both systems are still in their stochastic enrichment stage \citep{shetrone01,cohen10,tsujimoto17,frebel23}, which prevents conclusive interpretation regarding the evolution of [Eu/Mg], although \citet{duggan18} argued for an increasing \rpro\ trend in these galaxies, especially in Ursa Minor, using Ba measurements and an average trend in [Ba/Eu].
% Further complicating the picture, for Draco, it is found that the observed neutron-capture element abundances may have significant \spro\ contribution even at low metallicity \citep{cohen09,tsujimoto15,tsujimoto17}.
% For Ursa Minor, \citep{duggan18} argued the existence of delayed \rpro\ sources, NSMs, by studying Ba abundance evolution.

We also compare our results with measurements from the disrupted dwarf Wukong/LMS-1 stellar stream \citep{naidu20,yuan20b} by \citet{limberg23}.
A similar rise in [Eu/Mg] as found in the GSE is observed in Wukong/LMS-1, though we note that stars from this accreted system are only found up to [Mg/H]$\sim-1.2$. 
However, such a metallicity cutoff is expected given Wukong/LMS-1's lower estimated total mass, and hence, its lower star formation efficiency, when compared with GSE properties.  

Recent observations of the Large Magellanic Cloud (LMC) also show potential support for delayed \rpro\ sources \citep{reggiani21,chiti24}. 
For LMC stars at $\rm{[Fe/H]}>-2.5$, \citet{reggiani21} measured \rpro\ enhancement at the level of $\rm{[Eu/Fe]}\sim 0.7$, while \citet{chiti24} found no evidence of \rpro\ enhancement for LMC stars with $\rm{[Fe/H]}<-2.5$. 
The comparison suggests that there may be a similar rise present in the LMC, with the delayed \rpro\ sources changing the \rpro\ enhancement level at $\rm{[Fe/H]}\sim-2.5$. 

In summary, no other dwarf galaxies, except for tentative evidence from Wukong/LMS-1 and LMC, currently display any similarly clear and obvious rise in [Eu/Mg] with [Mg/H] as the GSE does. 
% An exception might be the recent measurements, although metallicity limited, in Wukong/LMS-1. 
Additional stellar data in the GSE and other systems will thus bring forward important clues about how the $r$-process rises within different galactic environments.

\section{Conclusions}
\label{sec:conclusion}

We study the \rpro\ elements in a sample of 43 GSE stars with a metallicity ranging between $\rm{[Fe/H]}\sim-2.5$ and $-1.0$.
Eleven metal-poor stars with $\rm{[Fe/H]}\lesssim -2.0$ allow us to probe the early \rpro\ enrichment, previously out of reach. 
We also derive Eu and Mg abundances to study the evolution of \rpro\ elements and their astrophysical sources in the GSE progenitor.

GSE is the second-ever dwarf galaxy for which we find a clear rise in [Eu/Mg].
Such a rise has been expected due to delayed \rpro\ sources, yet evidence has only recently come to the surface. 
% By mapping the [Mg/H] to stellar age, we find that this rise happened at $\sim13$-$12$\,Gyr ago. 
% This rise is clear evidence for the existence of delayed \rpro\ sources and can be used for estimating the delay time of NSMs.

We examine the rise of the \rpro\ in two scenarios where the increase of observed \rpro\ abundances may be explained either with (1) a short delay ($10$\,Myr) in NSMs and with low-yield rCCSNe producing minimal \rpro\ at early time, or with (2) a long delay ($500$\,Myr) in NSMs and with high-yield rCCSNe producing all \rpro\ elements at early time.
Both cases require delayed \rpro\ sources and predict effective Eu yields for different sites that are consistent with what is found in the literature.
Prompt sources alone can explain \rpro\ enrichment at early times but require yields close to the estimated Eu yield of a given NSM \citep{cote18}. 
If rCCSNe cannot supply enough \rpro\ element enrichment, NSMs must correspondingly have a shorter delay time. 
These takeaways are independent of many degeneracies typically encountered in modeling chemical enrichment processes in galaxies.

This study showcases the power of using massive disrupted dwarf galaxies in the Milky Way halo to constrain early \rpro\ enrichment and the astrophysical sites of the \rpro. 
Systems like the GSE have stellar populations that are relatively easily accessible for obtaining high-resolution spectroscopy of member stars, and they have a less complex merger history than the Milky Way that makes these kinds of explorations possible. 
Further constraining the delayed \rpro\ sources' delay time distribution and combined element yields will rely on additional observations that probe the more metal-poor population in the GSE as well as other accreted dwarf galaxies.

\begin{acknowledgements}
%xx acknowledge support from NSF grant xx.
X. O.  thanks the LSST Discovery Alliance Data Science Fellowship Program, which is funded by LSST Discovery Alliance, NSF Cybertraining Grant \#1829740, the Brinson Foundation, and the Moore Foundation; his participation in the program has benefited this work.
X. O. and A.F. acknowledge support from NSF grants AST-1716251 and AST-2307436.
A.P.J acknowledges support from NSF grants AST-2206264 and AST-2307599.
G.L. acknowledges FAPESP (procs. 2021/10429-0 and 2022/07301-5).

This work presents results from the European Space Agency (ESA) space mission \emph{Gaia}. \Gaia data are being processed by the \Gaia Data Processing and Analysis Consortium (DPAC). Funding for the DPAC is provided by national institutions, in particular the institutions participating in the Gaia MultiLateral Agreement (MLA). The Gaia mission website is \url{https://www.cosmos.esa.int/gaia}. The Gaia archive website is \url{https://archives.esac.esa.int/gaia}.

This research has made use of NASA's Astrophysics Data System Bibliographic Services; the arXiv pre-print server operated by Cornell University; the SIMBAD and VizieR databases hosted by the Strasbourg Astronomical Data Center

\end{acknowledgements}

\software{%
\texttt{matplotlib} \citep{hunter07},
\texttt{numpy} \citep{vanderwalt11},
\texttt{scipy} \citep{jones01},
\texttt{astropy} \citep{2013A&A...558A..33A,2018AJ....156..123A}, and
\texttt{SMHR} \citep{casey14}.}

\clearpage

\bibliographystyle{aasjournal}
\bibliography{xou}

\setcounter{equation}{0}
\setcounter{figure}{0}
\setcounter{table}{0}
\setcounter{section}{0}
\makeatletter
\renewcommand{\theequation}{S\arabic{equation}}
\renewcommand{\thefigure}{S\arabic{figure}}
\renewcommand{\thetable}{S\arabic{table}}

% \appendix 

% \section{Appendix A}
% \label{app:appendixA}

\end{CJK*}
\end{document}